\documentclass[twocolumn]{autart}    

\usepackage{cite}
\usepackage{amsmath,amssymb,amsfonts}
\usepackage{algorithm}
\usepackage{algorithmicx}
\usepackage{algpseudocode}
\usepackage{graphicx}
\usepackage{mathrsfs}
\usepackage{textcomp}
\usepackage{xcolor}
\usepackage{caption}
\usepackage{multirow} 
\usepackage{array,booktabs}

\newtheorem{theorem}{Theorem}
\newtheorem{definition}{Definition} 
\newtheorem{lemma}{Lemma} 
\newtheorem{remark}{Remark}

\newtheorem{proposition}{Proposition}
\newtheorem{assumption}{Assumption}

\begin{document}

\begin{frontmatter}

\title{Integrating Uncertainties for Koopman-Based Stabilization \thanksref{footnoteinfo}} 

\thanks[footnoteinfo]{This paper was not presented at any IFAC 
	meeting. Corresponding author Z.~Duan.}

\author[SAMR]{Yicheng Lin}\ead{linyc020709@stu.pku.edu.cn},
\author[SAMR]{Bingxian Wu}\ead{davidwu2003@stu.pku.edu.cn},
\author[HK]{Nan Bai}\ead{eenanbai@ust.hk},
\author[SAMR]{Zhiyong Sun}\ead{zhiyong.sun@pku.edu.cn},
\author[COE]{Yunxiao Ren}\ead{renyx@pku.edu.cn},
\author[YP]{Chuanze Chen}\ead{chenchuanze@stu.pku.edu.cn},
\author[SAMR]{Zhisheng Duan}\ead{duanzs@pku.edu.cn}

\address[SAMR]{School of Advanced Manufacturing and Robotics, Peking University, Beijing}
\address[HK]{Department of Electronic and Computer Engineering, The Hong Kong University of Science and Technology, Hong Kong}
\address[COE]{College of Engineering, Peking University, Beijing}
\address[YP]{Yuanpei College, Peking University, Beijing}
          

\begin{abstract}                          
Over the past decades, the Koopman operator has been widely applied in data-driven control, yet its theoretical foundations remain underexplored. This paper establishes a unified framework to address the robust stabilization problem in data-driven control via the Koopman operator, fully accounting for three uncertainties: projection error, estimation error, and process disturbance. It comprehensively investigates both direct and indirect data-driven control approaches, facilitating flexible methodology selection for analysis and control. For the direct approach, considering process disturbances, the lifted-state feedback controller, designed via a linear matrix inequality (LMI), robustly stabilizes all lifted bilinear systems consistent with noisy data. For the indirect approach requiring system identification, the feedback controller, designed using a nonlinear matrix inequality convertible to an LMI, ensures closed-loop stability under worst-case process disturbances. Numerical simulations via cross-validation validate the effectiveness of both approaches, highlighting their theoretical significance and practical utility.
\end{abstract}

\end{frontmatter}

\section{Introduction}
\subsection{Koopman Operator and System Transformations}
Control theory of nonlinear systems is more complicated than that of linear ones and has attracted increasing interests owing to its significance. For decades, researchers have been considering to provide systematic approach to analyze and control nonlinear systems. With this thought, the Koopman operator, originated from B.O.Koopman’s work in 1931 about Hamiltonian systems \cite{koopman1931hamiltonian}, has gained growing popularity. By describing dynamics on the space of observables, the linear operator is able to transform finite-dimensional nonlinear autonomous systems (without control) into infinite-dimensional linear ones.

Research has demonstrated that for control-affine nonlinear systems, the Koopman operator enables transformation into bilinear form under mild assumptions \cite{goswami2021bilinearization}. Nevertheless, prevalent methodological limitations persist within this research domain. Specifically, a large body of works focusing on controller design across practical scenarios has inappropriately assumed that nonautonomous nonlinear systems can be converted into fully linear form via the Koopman operator \cite{TSMC2024Resilient},\cite{YvTRO2024Autogeneration}. While the designed controllers have been validated through numerical simulations and even physical experiments, the absence of rigorous analysis regarding approximation errors and their implications for closed-loop stability guarantees render such results theoretically questionable. Conversly, although recent advances have explored bilinear system embeddings for control-affine dynamics \cite{2019Data},\cite{9837986}, a large amount of them neglect the impact of uncertainties. This oversight potentially undermines robustness and limits the performance of these methods.

\subsection{Koopman Operator and Data-Driven Control}
Undoubtedly anyway, the Koopman operator has been successfully used for stability analysis \cite{mauroy2016global},\cite{yi2023equivalence}, system identification \cite{CDC2016observer},\cite{goswami2021bilinearization}, optimal control \cite{villanueva2021towards},\cite{TAC2025KoopmanHJB}, etc. With the growing availability of numerous data and especially the emergence of (extended) dynamic mode decomposition (DMD/EDMD) algorithm to approximate the Koopman operator \cite{2017On}, people have been exploring its potential in both data-driven control theory and method \cite{8619727},\cite{9867811}, as well as applications in fluid mechanics, power grids, unmanned systems \cite{mauroy2020koopman},\cite{manzoor2023vehicular},\cite{Budi2012Applied}, etc.

Stability remains a fundamental requirement for data-driven control via the Koopman operator. Existing studies have endeavored to develop a general framework or methodology for addressing feedback stabilization in this domain. Early works made beneficial attempts using control Lyapunov functions (CLFs), but confined controllers to specific forms - such as piecewise constant, quadratic, and modified Sontag’s formulas \cite{8619727, mauroy2020koopman}. Concurrently, linear matrix inequality (LMI)-based approaches leveraging Lyapunov functions were proposed for feedback design \cite{9837986}. Notably, these early efforts overlooked the impact of uncertainties on stabilization.

Data-driven control approaches are categorized into indirect and direct methods, with the key distinction being whether system identification is involved. For indirect Koopman-based data-driven control methods, the canonical workflow involves selecting a set of dictionary functions, collecting and preprocessing data, approximating the Koopman operator via (E)DMD, identifying the lifted system, and finally designing the controller. In contrast, direct methods bypass the identification step and design the controllers from data directly. Three primary sources of unavoidable uncertainties emerge in this context:
\begin{itemize}
	\item Projection error arising from finite dictionary functions. To fully characterize nonlinear dynamics, the Koopman operator typically requires lifting the original system to a high-dimensional even infinite-dimensional space. In practice, however, truncation to lower dimensions is necessary.
	\item Estimation error due to finite data collection. To accurately characterize the nonlinear dynamics, sufficient data are usually demanded. This is a great challenge in most cases, resulting in incapability of exact knowledge of underlying nonlinear systems.
	\item Noise or disturbance of collected data. In ideal numerical simulation, we are able to obtain accurate enough data thanks to the advanced computer technology. But noise-free and disturbance-free precise data hardly exist in practice, which inevitably affects the robustness even performance of designed controllers.
\end{itemize}
Unless otherwise specified, the first two categories will be referred to as \textit{approximation error} in what follows.

The widespread applicability and profound impact of the Koopman operator in data-driven control underscore the need for a more robust theoretical foundation. Given that the aforementioned uncertainties cannot be fully eliminated in practice, developing a general framework to systematically analyze their effects on stabilization is imperative—this would help mitigate their impacts in control engineering applications. Unfortunately, limited studies have addressed this direction. For Koopman-based direct data-driven control, to our knowledge, only \cite{CDC2022You} has examined stabilization. For indirect approaches, \cite{Strasser2023Robust},\cite{Strasser2024Koopman} conducted preliminary explorations, analyzing the impact of probabilistic estimation error and ensuring robustness in controller design. By explicitly accounting for such error, these works address a critical limitation of earlier studies: they focus on stabilizing the original nonlinear system, rather than the transformed bilinear one via the Koopman operator. 

Nevertheless, existing works suffer from significant limitations that critically hinder the advancement and practical adoption of Koopman-based data-driven control. Two critical gaps stand out:
\begin{itemize}
	\item Unnecessary restrictive constraints undermine the practical viability of existing results. For instance, \cite{CDC2022You} introduced excessive conservatism in handling bilinear terms (detailed in Section 5.2), yielding overly conservative controller designs. Similarly, predefined constraint set for an extra uncertainty variable in \cite{Strasser2024Koopman}, which is hard to define with scarce prior system knowledge, directly limits the method's applicability.
	
	\item Current literature ignores critical uncertainties, endangering the reliability of controllers. The forward-invariance assumption on dictionary functions \cite[Assumption 2]{Strasser2024Koopman}, though analytically convenient, is rarely verifiable and obscure the impact of projection error, weakening the rigor of stability guarantees. Moreover, unaddressed noise or disturbance \cite{CDC2022You, Strasser2023Robust, Strasser2024Koopman} may lead to performance degradation or instability in practice.
\end{itemize}

\subsection{Motivations and Contributions}
Building upon the preceding discussions, this paper aims to simultaneously handle the aforementioned three categories of uncertainties in a unified framework to address nonlinear feedback stabilization problems in Koopman-based data-driven control. Our theoretical contributions deliver dual-functional values, not only in practically synthesizing robust controllers, but also in theoretically analyzing effects of different uncertainties and evaluating the robustness of pre-designed controllers. This ensures the generality of proposed results in both theory and application. Additionally with this thought in mind, direct and indirect data-driven approaches are investigated respectively, enabling a flexible selection of appropriate methodologies for analysis and control. As a significant tool of robust control theory, Petersen's lemma is fully utilized \cite{BISOFFI2022Petersen} to derive the theoretical results.

The main contributions are listed as follows.
\begin{itemize}
	\item We account for unknown-but-bounded process disturbance (with a similar approach applicable to bounded noise) and integrate it with the approximation error into a unified framework, enabling systematic analysis of their impacts on Koopman-based nonlinear stabilization. To our knowledge, this constitutes the first comprehensive investigation of all such uncertainties, yielding general theoretical results for closed-loop stability guarantees.
	
	\item For direct data-driven control approach, an LMI-based lifted-state feedback controller design method is proposed to stabilize all of the bilinear systems \textit{consistent with noisy data} even in the presence of approximation error. This addresses a critical gap in existing literature on Koopman-based direct data-driven control, also serves as a criterion to analyze the robustness of pre-designed controllers as well.
	
	\item For indirect data-driven control approach, we propose a lifted-state feedback controller based on a concise nonlinear matrix inequality, which extends the findings of \cite{Strasser2024Koopman} by explicitly incorporating the process disturbance (noise) alongside the projection error. This nonlinear matrix inequality can be converted into an LMI for computation and also serves as a criterion to analyze the robustness of pre-designed controllers.
\end{itemize}

This paper is organized as follows. Section~2 recalls some basic mathematics mainly involving the Koopman operator and Petersen's lemma. Sections 3 and 4 discuss direct and indirect data-driven control approach respectively, followed by some discussions in Section~5. Section~6 demonstrates the proposed results via numerical examples with conclusions in Section~7.

\textbf{Notations: }Throughout this paper, we write $I_p$ for the $p\times p$ identity matrix and $0_{p\times q}$ for the $p\times q$ zero matrix, where we omit the index if the dimension is clear. Matrix blocks which can be inferred from symmetry are denoted by $*$, and without causing ambiguity, we abbreviate $B^\top AB$ or $A+A^\top$ by writing $[*]^\top AB$ or $A+[*]^\top$ somewhere. $A^\dagger$ denotes the Moore-Penrose pseudo-inverse of $A$. Other notations are ordinary.
\section{Mathematical Preliminaries}
\subsection{The Koopman Operator and Nonlinear System Reformulation}
We start with the unactuated nonlinear system
\begin{equation}\label{unactuated system}
	\dot{x}=f\left(x\right)
\end{equation}
where $x\in\mathbb{X}\subseteq\mathbb{R}^n$ denotes the state and $f(0)=0$, i.e., the origin is an equilibrium of the unactuated system. We define a Banach space $\mathcal{F}$ of observables $\varphi:\mathbb{X}\rightarrow\mathbb{R}$, and the Koopman operator is defined as follows.

\begin{definition}
	The continuous time Koopman operator $\mathcal{K}^t:\mathcal{F} \rightarrow \mathcal{F}$ is defined as
	\begin{equation}
		(\mathcal{K}^t\varphi)(x_0)=\varphi\circ S(t,x_0) \label{operator definition}
	\end{equation}
	where $\circ$ denotes the function composition and $S(t,x_0)$ denotes the flow map (solution) of system \eqref{unactuated system} at time $t>0$ with an initial state $x_0$. Furthermore, under the assumption that $\varphi(x)$ is continuously differentiable, it satisfies
	\begin{equation}
		\frac{\mathrm{d}\varphi(x)}{\mathrm{d}t}=\mathcal{L}_f\varphi\triangleq\lim_{t\to0}\frac{(\mathcal{K}^t\varphi-\varphi)}{t}=\nabla\varphi\cdot f \label{generator definition}
	\end{equation}
	where $\mathcal{L}_f$ is defined as the infinitesimal generator, also equals to the Lie derivative with respect to $f$.
\end{definition}

We note that $\mathcal{K}^t(\alpha \varphi_1+\beta \varphi_2)=\alpha \mathcal{K}^t\varphi_1 +\beta \mathcal{K}^t\varphi_2$, which indicates the linearity of Koopman operator even if the dynamics is nonlinear.

\begin{definition}
	A Koopman eigenfunction corresponding to the unactuated system \eqref{unactuated system} is an observable $\phi_\lambda\in\mathcal{F}$ such that
	\begin{equation}
		\mathcal{K}^t\phi_\lambda=e^{\lambda t}\phi_\lambda
	\end{equation}
	for some $\lambda\in\mathbb{C}$, which is the associated Koopman eigenvalue. Additionally with \eqref{generator definition}, we obtain
	\begin{equation}
		\mathcal{L}_f\phi_\lambda=\nabla\phi_\lambda\cdot f=\lambda\phi_\lambda. \label{eigenfunction definition generator}
	\end{equation}
\end{definition}
It is known that if $\phi_{\lambda_1}$ and $\phi_{\lambda_2}$ are Koopman eigenfunctions with eigenvalues $\lambda_1$ and $\lambda_2$ respectively, $\phi_{\lambda_1}^{k_1}\phi_{\lambda_2}^{k_2}$ is also an eigenfunction with eigenvalue $k_1\lambda_1+k_2\lambda_2$, which means there are perhaps infinitely many eigenfunctions.

The Koopman operator provides a linear representation framework for nonlinear dynamical systems. To formalize this, we define a finite-dimensional dictionary of observables $\Psi(x)=[\psi_1(x)\ \cdots \ \psi_N(x)]^\top\in\mathbb{R}^N$ with each of $\psi_i:\mathbb{X}\rightarrow\mathbb{R}$ continuously differentiable. In general, to fully capture nonlinear dynamics, embedding dimension $N$ usually exceeds that of original state $n$ (i.e., $N>n$), leading to the concept of lifted system representation.

When these observables form Koopman-invariant subspaces, they enable exact linear embeddings of \eqref{unactuated system}. An important case emerges when the dictionary consists of Koopman eigenfunctions $\Phi(x)=[\phi_{\lambda_1}(x)\ \cdots \ \phi_{\lambda_N}(x)]^\top$ associated with eigenvalues $\{\lambda_i\}_{i=1}^N$, inducing
\begin{equation}
	\frac{\mathrm{d}}{\mathrm{d}t}\Phi(x) = \Lambda\Phi(x), \quad \Lambda = \operatorname{diag}(\lambda_1,\cdots,\lambda_N).
\end{equation}
More generally, any dictionary $\Psi(x)$ related to eigenfunctions through invertible linear transformation $T\in \mathbb{R}^{N\times N}$ (i.e., $\Psi(x)=T\Phi(x),\ \det(T)\neq0$) preserves linearity in the lifted coordinates.

Certainly, beyond Koopman eigenfunctions, dictionary functions can be selected via diverse methods, such as various polynomials \cite{TSMC2024Resilient} or kernels \cite{strasser2025kernelbasederrorboundsbilinear}. A general approach still remains challenging.

Now we consider the control-affine nonlinear system with \textit{unknown-but-bounded} process disturbance, i.e.,
\begin{equation}\label{original system}
	\dot{x}(t)=f\left(x\right)+\sum_{i=1}^{m}g_i\left(x\right)u_i+\overline{d} 
\end{equation}
where $x\in \mathbb{X}, u=\left[u_1\ \cdots\ u_m\right]^\top\in\mathbb{U}$ are the state and input respectively. The disturbance $\overline{d}$ can be a function of $t,x$. $\mathbb{X}\subseteq\mathbb{R}^{n}$ and $\mathbb{U}\subseteq\mathbb{R}^m$ are assumed as compact sets, also $f$, $g_i,i=1,\cdots,m$ are assumed to be continuously differentiable. This paper aims to design a proper controller $\kappa:\mathbb{R}^n\rightarrow\mathbb{R}^m$ such that the origin of closed-loop system \eqref{closed-loop system} is stable under a bounded process disturbance $\overline{d}$ in whatever possible form.
\begin{equation}\label{closed-loop system}
	\dot{x}(t)=f\left(x\right)+\sum_{i=1}^{m}g_i\left(x\right)\kappa_i\left(x\right)+\overline{d}
\end{equation}
Now we transform the control-affine nonlinear system via the Koopman operator. The time derivative of $\Psi$ is
\begin{equation}\label{lifted system operator}
	\begin{aligned}
		\frac{\mathrm{d}\Psi}{\mathrm{d}t}&=\nabla\Psi\cdot f(x)+\sum_{i=1}^m u_i\nabla\Psi\cdot g_i(x)+\nabla\Psi\cdot \overline{d} \\
		&=\mathcal{L}_f\Psi+\sum_{i=1}^mu_i\mathcal{L}_{g_i}\Psi+d.
	\end{aligned}
\end{equation}
We further assume $\Psi(0)=0$, $\Psi^{-1}(0)\cap\mathbb{X}=\{0\}$ and $\exists C\in\mathbb{R}^{n\times N}, \text{s.t. }x=C\Psi(x)$ to facilitate Lyapunov based stability analysis. As shown in \cite{goswami2021bilinearization}, if the eigenspace of $\mathcal{L}_f$ is an invariant subspace of $\mathcal{L}_{g_i},i=1,\cdots,m$, system \eqref{lifted system operator} (hence \eqref{original system}) without process disturbance is bilinearizable. If this is not satisfied, a projection error term can be introduced such that
\begin{equation}\label{lifted system AB}
	\frac{\mathrm{d}\Psi(x)}{\mathrm{d}t}=A\Psi(x)+B_0u+\sum_{i=1}^m u_iB_i\Psi(x)+r(\Psi(x),u) +d.
\end{equation}
In data-driven control problem where $A,B_0$ and $B_i,i=1,\cdots,m$ are obtained from data, not only projection error but also estimation error resulting from finite data should be considered. The following lemma gives a proportional error bound on the approximation error.
\begin{lemma}
	Suppose the collected data are independent and identically distributed (i.i.d.) in $\mathbb{X}$. Then, there exist constants $c_1,c_2$ such that the approximation error term in \eqref{lifted system AB} including projection error and estimation error is bounded by \eqref{error bound} for all $(x,u)\in\mathbb{X}\times\mathbb{U}$.
	\begin{equation}\label{error bound}
		\|r(\Psi(x),u)\|\leq c_1\|\Psi(x)\|+c_2\|u\|
	\end{equation}
\end{lemma}
Here we note that the approximation error bound has been investigated in several recent papers. For instance, \cite[Proposition 5]{Strasser2024Koopman} derived a probabilistic bound for the estimation error $c_1,c_2\in \mathcal{O}(1/\sqrt{\delta d_0})$, where $\mathcal{O}(\cdot)$ is the Big O notation, characterizing the asymptotic upper bound of a function, $\delta$ and $d_0$ denote the probability tolerance and data amount respectively. Leveraging kernel-based methods, deterministic bound for the approximation error was established in \cite[Theorem 5]{strasser2025kernelbasederrorboundsbilinear}. These results ensure the proportional error relationship \eqref{error bound} vanishing at desired origin $(x,u)=(0,0)$, which is crucial for robust stabilization \cite{Strasser2025KoopmanGuarantee} of the original nonlinear control-affine system.

\subsection{Petersen's Lemma}
Since we investigate the influences of approximation error and disturbances on feedback stabilization, necessary tools of robust control should be introduced. In this paper, we advocate the use of Petersen's lemma \cite{BISOFFI2022Petersen}.
\begin{lemma}
	Consider matrices $\mathbf{C}\in\mathbb{R}^{n\times n}$, $\mathbf{M}\in\mathbb{R}^{n\times p}$, $\overline{\mathbf{D}}\in\mathbb{R}^{q\times q}$, $\mathbf{N}\in\mathbb{R}^{q\times n}$ with $\mathbf{C}=\mathbf{C}^\top$ and $\overline{\mathbf{D}}=\overline{\mathbf{D}}^\top\succeq 0$. Let the set of matrices
	\begin{equation}\label{setDdefinition}
		\mathcal{D}:=\{\mathbf{D}\in\mathbb{R}^{p\times q}:\mathbf{D}^\top\mathbf{D}\preceq \overline{\mathbf{D}}\}.
	\end{equation}
	Then, 
	\begin{equation}
		\mathbf{C}+\mathbf{M}\mathbf{D}\mathbf{N}^\top+\mathbf{N}\mathbf{D}^\top\mathbf{M}^\top\prec 0,\ \forall \mathbf{D}\in\mathcal{D}
	\end{equation}
	if and only if there exists $\mu> 0$ such that 
	\begin{equation}
		\mathbf{C}+\mu\mathbf{M}\mathbf{M}^\top+\mu^{-1}\mathbf{N}\overline{\mathbf{D}}\mathbf{N}^\top\prec 0.
	\end{equation}
\end{lemma}
Petersen's lemma serves as an ideal approach for matrix elimination \cite{BISOFFI2022Petersen}, since we need not to verify the expected inequality for all matrices satisfying the given norm bound. This explains why it can become a useful tool for robust control. Also we have a corresponding non-strict version, given in Lemma 3.
\begin{lemma}
	Consider matrices $\mathbf{C}\in\mathbb{R}^{n\times n}$, $\mathbf{M}\in\mathbb{R}^{n\times p}$, $\overline{\mathbf{D}}\in\mathbb{R}^{q\times q}$, $\mathbf{N}\in\mathbb{R}^{q\times n}$ with $\mathbf{C}=\mathbf{C}^\top$ and $\overline{\mathbf{D}}=\overline{\mathbf{D}}^\top\succeq 0$. Let the set $\mathcal{D}$ be defined the same as \eqref{setDdefinition}, and assume $\mathbf{M}\neq 0,\overline{\mathbf{D}}\succ 0$ and $\mathbf{N}\neq 0$. Then, 
	\begin{equation}
		\mathbf{C}+\mathbf{M}\mathbf{D}\mathbf{N}^\top+\mathbf{N}\mathbf{D}^\top\mathbf{M}^\top\preceq 0,\ \forall \mathbf{D}\in\mathcal{D}
	\end{equation}
	if and only if there exists $\mu> 0$ such that 
	\begin{equation}
		\mathbf{C}+\mu\mathbf{M}\mathbf{M}^\top+\mu^{-1}\mathbf{N}\overline{\mathbf{D}}\mathbf{N}^\top\preceq 0.
	\end{equation}
\end{lemma}
The proofs of Lemmas 2 and 3 can be found in \cite{BISOFFI2022Petersen}, and are therefore omitted here due to space constraints.


\section{Direct Data-Driven Control Approach}
Building on the Koopman operator theory, this section addresses the robust stabilization problem of nonlinear control-affine systems in direct data-driven control. Uncertainties of different forms, the approximation error and process disturbance, are considered in a stepwise manner. Given the unknown disturbance, a set of bilinear systems will be consistent with the noisy data. Ensuring that the designed controller can stabilize these systems even in the presence of approximation error will guarantee the closed-loop stability.
\subsection{Problem Formulation}
For the lifted bilinear system via the Koopman operator,
\begin{equation}\label{bilinear system z disturbance}
	\dfrac{\mathrm{d}z}{\mathrm{d}t}=Az+B_0u+\sum_{i=1}^mu_iB_iz+r(z,u)+d
\end{equation}
where we denote $z=\Psi(x)$ for simplicity (actually \eqref{bilinear system z disturbance} is equivalent to \eqref{lifted system operator}), the noisy data can be collected,
\begin{equation}\label{data set}
	\begin{aligned}
		U_0:&=\left[u_0 \quad u_1 \quad \cdots \quad u_{T-1}\right] \\
		Z_0:&=\left[z_0 \quad z_1 \quad \cdots \quad z_{T-1}\right] \\
		&=\left[\Psi(x_0) \quad \Psi(x_1) \quad \cdots \quad \Psi(x_{T-1})\right] \\
		V_0^i:&=\left[u_0^iz_0 \quad u_1^iz_1 \quad \cdots \quad u_{T-1}^iz_{T-1}\right] \\
		Z_1:&=\left[\dot{z}_0 \quad \dot{z}_1 \quad \cdots \quad \dot{z}_{T-1}\right]
	\end{aligned}
\end{equation}
where $u^i_j,\ i=1,\cdots,m,\ j=0,1,\cdots,T-1$ denotes the $i$-th component of the control input $u_j$.
\begin{remark}
	In this paper, we assume that the derivative information $\dot{x}_j, j=0,1,\cdots,T-1$ can be collected, which is a standard requirement in continuous-time data-driven control \cite{Strasser2024Koopman, BISOFFI2022Petersen}. This assumption also leads to the knowledge of $\dot{z}_j=\nabla\Psi(x_j)\cdot \dot{x}_j$. There are also established techniques for derivative estimation under noisy and sampled-data scenarios, e.g. \cite{Rapisarda2024Orthogonal}. There is no assumption on the method of collecting data, which means that $Z_0$ can be collected either along a single trajectory or from multiple trajectories.
\end{remark}

Since the approximation error can be handled by $r(z,u)$, we temporarily discard the error term. Although the disturbance is unknown, we can also arrange the sequence
\begin{equation}\label{definition D0}
	\begin{aligned}
		D_0:&=[d_0 \ d_1 \ \cdots \ d_{T-1}] \\
		&=[\nabla\Psi(x_0) \overline{d}_0 \ \nabla\Psi(x_1)\overline{d}_1 \ \cdots \ \nabla\Psi(x_{T-1}) \overline{d}_{T-1}],
	\end{aligned}
\end{equation}
such that the collected data should satisfy
\begin{equation}\label{Relations ZUV}
	Z_1=AZ_0+B_0U_0+\sum_{i=1}^mB_iV_0^i+D_0.
\end{equation}

Since the process disturbance has bounded energy and the considered state space is a compact set $\mathbb{X}$, we have the following assumption without loss of generality.
\begin{assumption}
	Disturbance sequence $D_0$ is of bounded energy, i.e., $D_0\in\mathcal{D}$ such that for some matrix $\Delta$,
	\begin{equation}\label{Assumption 1}
		\mathcal{D}:=\left\{D\in\mathbb{R}^{n\times T}:DD^\top\preceq\Delta\Delta^\top\right\}.
	\end{equation}
\end{assumption}
Then, we introduce an important concept for the set $\mathcal{C}$ of matrices, called \textit{consistency with data}, i.e.,
\begin{equation}
	\begin{aligned}
		\mathcal{C}:&=\big\{[A\ B_0\ B_1\ \cdots B_m]: \\
		&Z_1=AZ_0+B_0U_0+\sum_{i=1}^mB_iV_0^i+D_0, D_0\in\mathcal{D}\big\}.
	\end{aligned}
\end{equation}
To address the robust stabilization problem, it is necessary to reformulate the set $\mathcal{C}$. The fundamental ideas draw on \cite{BISOFFI2022Petersen}, with adaptations tailored to bilinear systems. The set $\mathcal{C}$ is equivalently reformulated as
\begin{equation}
	\begin{aligned}
		\mathcal{C}=\big\{&\mathbf{Z}^\top:=[A\ B_0\ B_1\ \cdots B_m]: \\
		&\mathbf{Z}^\top\mathbf{A}\mathbf{Z}+\mathbf{Z}^\top\mathbf{B}+\mathbf{B}^\top\mathbf{Z}+\mathbf{C}\preceq 0\big\},
	\end{aligned}
\end{equation}
where 
\begin{equation}\label{DataMatABC}
	\left[
	\begin{array}{c|c}
		\mathbf{C} & \mathbf{B}^\top \\
		\hline
		\mathbf{B} & \mathbf{A}
	\end{array}
	\right]=
	\left[
	\begin{array}{c|c}
		Z_1 Z_1^\top - \Delta \Delta^\top & -Z_1 W_0^\top \\
		\hline
		-W_0 Z_1^\top & W_0 W_0^\top
	\end{array}
	\right], W_0=
	\begin{bmatrix}
		Z_0 \\ U_0 \\V_0^1 \\ \vdots \\V_0^m
	\end{bmatrix}.
\end{equation}

An assumption on the matrix $W_0$ in \eqref{DataMatABC} is imposed as below.
\begin{assumption}
	The matrix $W_0$ is of full row rank.
\end{assumption}
Then, the set $\mathcal{C}$ is further reformulated in a compact form as 
\begin{equation}\label{Roformulation zetaQ}
	\begin{aligned}
		\mathcal{C}=\big\{\mathbf{Z}^\top:
		(\mathbf{Z}-\boldsymbol{\zeta})^\top\mathbf{A}(\mathbf{Z}-\boldsymbol{\zeta})\preceq \mathbf{Q}\big\},
	\end{aligned}
\end{equation}
where, by $\mathbf{A}\succ 0$, we define
\begin{equation}\label{definition of zeta and Q}
	\boldsymbol{\zeta}:=-\mathbf{A}^{-1}\mathbf{B},\quad \mathbf{Q}:=\mathbf{B}^\top \mathbf{A}^{-1}\mathbf{B}-\mathbf{C}.
\end{equation}
Equation \eqref{Roformulation zetaQ} is well-defined, since following similar thoughts to \cite[Lemma 1]{BISOFFI2022Petersen}, we can prove that $\mathbf{A}\succ 0, \mathbf{Q}\succeq 0$ under Assumption 2. This also leads to a final reformulation of the set $\mathcal{C}$, i.e.,
\begin{equation}\label{C=E reformulation}
	\mathcal{C}=\mathcal{E}:= \left\{ \bigl( \boldsymbol{\zeta} + \mathbf{A}^{-1/2} \boldsymbol{\gamma} \mathbf{Q}^{1/2} \bigr)^\top : \| \boldsymbol{\gamma} \| \leq 1 \right\}.
\end{equation}

\subsection{Main Results for Direct Data-Driven Control}
In order to address the approximation error $r(z,u)$, the following proposition should be introduced.
\begin{proposition}
    Consider the lifted bilinear system via the Koopman operator with no disturbance, i.e.,
    \begin{equation}\label{bilinear system no disturbance}
	\dfrac{\mathrm{d}z}{\mathrm{d}t}=Az+B_0u+\sum_{i=1}^mu_iB_iz+r(z,u)
    \end{equation}
    where the overall approximation error satisfies Lemma 1. The lifted-state feedback controller $u=Kz=L^\top P^{-1}z$ achieves exponential stability of \eqref{original system} without disturbance with guaranteed region of attraction $\mathbb{X}_{RoA}=\{x\in \mathbb{X}:z^\top P^{-1}z=\Psi(x)^\top P^{-1}\Psi(x)\leq c\}$, if there exist $L\in\mathbb{R}^{N\times m},0\prec P=P^\top\in \mathbb{R}^{N\times N}$ and scalars $\alpha,\beta,\nu,\tau$ with $\frac{1}{\alpha} +\frac{1}{\beta}=1$, such that
	\begin{equation}\label{NMI}
		\begin{aligned}
			&PA^\top+B_0L^\top+AP+LB_0^\top+\nu\sum_{i=1}^{m} B_iPB_i^\top \\ +&\nu^{-1}cLL^\top+\tau^{-1}(\alpha c_1^2P^2+\beta c_2^2LL^\top)+\tau I_{N}\prec 0.
		\end{aligned}
	\end{equation}
\end{proposition}
In this paper, Proposition 1 serves as a closed-loop stability criterion under the effect of approximation error. The proof is provided in Appendix A for completeness. Since the controller ensures exponential stability of \eqref{original system} in disturbance-free scenarios ($d=0$), it consequently induces input-to-state stability (ISS) with respect to bounded disturbances, as per \cite[Lemma 9.2]{khalil1996nonlinear}.

Notably, Proposition 1 does not require \textit{a priori} fixed set for the additional uncertainty variable $\Delta_\Phi$ as introduced in \cite{Strasser2024Koopman}, leaving only a single matrix inequality \eqref{NMI} to solve that can be converted to an LMI. Parameters $\alpha,\beta$ can be flexibly adjusted via error coefficients $c_1,c_2$. Particularly, setting $\alpha=\beta=2, c=1$ reduces to the case \cite[Theorem 6]{Strasser2024Koopman}. Thus, Proposition 1 provides new insights while simplifying controller design.

A core idea in this paper is that three types of uncertainties should be addressed via distinct methods based on their inherent properties. Following \cite{Strasser2024Koopman}, estimation error can be bounded by $\|z\|$ and $\|u\|$. Projection error, expressed via \eqref{lifted system operator} and \eqref{bilinear system z disturbance} as
\begin{equation}
    r_p(z,u)=\mathcal{L}_fz-Az+\sum_{i=1}^mu_i\left(\mathcal{L}_{g_i}z-B_{0,i}-B_iz\right),
\end{equation}
can also be bounded with $\|z\|$ and $\|u\|$ (cf.\cite{Strasser2025KoopmanGuarantee},\cite{strasser2025kernelbasederrorboundsbilinear}). By contrast, the unstructured nature of unknown disturbance $d$ renders such bounding infeasible, necessitating the introduction of $\mathcal{C}$ to encode all bilinear systems consistent with noisy data. This enables the integration of diverse uncertainties into a unified framework.

Now we give the main result for direct data-driven stabilization problem. Without loss of generality, we consider the single-input case $m=1$, and the multi-input case is a natural extension.
\begin{theorem}
	Assume the data sets $Z_0,U_0,V_0^1$ collected from disturbed system \eqref{bilinear system z disturbance} (or equivalently \eqref{original system}) satisfy Assumption 2. The lifted-state feedback controller $u=L^\top P^{-1}z$ ensures closed-loop stability for \eqref{original system} with respect to any possible disturbance, if there exist $0\prec P=P^\top\in\mathbb{R}^{N\times N}, L\in\mathbb{R}^{N\times 1}$, and positive scalars $\mu,\alpha,\beta,\tau,\nu$ where $\frac{1}{\alpha}+\frac{1}{\beta}=1$, such that the matrix
	\begin{equation}\label{Direct LMI}
		\left[
		\begin{array}{c|c}
			\begin{matrix}
				-\frac{1}{\mu}\mathbf{C}+\tau I_N& 0& \left(\begin{bmatrix}
					P \\ L^\top \\ 0
				\end{bmatrix}-\frac{1}{\mu}\mathbf{B}\right)^\top \\
				0& -\frac{1}{\nu}P& \begin{bmatrix}
					0 \\ 0 \\ P
				\end{bmatrix}^\top \\
				\left(\begin{bmatrix}
					P \\ L^\top \\ 0
				\end{bmatrix}-\frac{1}{\mu}\mathbf{B}\right)& -\begin{bmatrix}
					0 \\ 0 \\ P
				\end{bmatrix}& -\frac{1}{\mu} \mathbf{A}
			\end{matrix}& E_{12} \\ \hline
			E_{12}^\top& E_{22}			
		\end{array}
		\right]
	\end{equation}
	is negative definite, where other blocks are
	\begin{equation}\label{Definition E12E22}
		E_{12}=
		\begin{bmatrix}
			L& P& L \\
			0& 0& 0 \\
			0& 0& 0 \\
		\end{bmatrix},E_{22}=
		\begin{bmatrix}
			-\frac{\nu}{c}& 0& 0 \\
			0& -\frac{\tau}{\alpha c_1^2}I_N& 0 \\
			0& 0& -\frac{\tau}{\beta c_2^2}
		\end{bmatrix}.
	\end{equation}
	Moreover, the region of attraction contains the set $\mathbb{X}_{RoA}=\{x\in \mathbb{X}:z^\top P^{-1}z=\Psi(x)^\top P^{-1}\Psi(x)\leq c\}$.
\end{theorem}
\begin{pf}
	By Schur complement lemma, the negative definiteness of \eqref{Direct LMI} is equivalent to $E_{22}\prec 0$ and
	\begin{equation}
		\begin{bmatrix}
			\begin{matrix}
				-\frac{1}{\mu}\mathbf{C}+\tau I_N+\frac{\alpha c_1^2}{\tau}P^2 \\
				+(\frac{\beta c_2^2}{\tau}+\frac{c}{\nu})LL^\top
			\end{matrix}& 0& \begin{bmatrix}
				P \\ L^\top \\ 0
			\end{bmatrix}^\top-\frac{1}{\mu}\mathbf{B}^\top \\
			0& -\frac{1}{\nu}P& -\begin{bmatrix}
				0 \\ 0 \\ P
			\end{bmatrix}^\top \\
			\begin{bmatrix}
				P \\ L^\top \\ 0
			\end{bmatrix}-\frac{1}{\mu}\mathbf{B}& -\begin{bmatrix}
				0 \\ 0 \\ P
			\end{bmatrix}& -\frac{1}{\mu}\mathbf{A}
		\end{bmatrix}\prec 0.
	\end{equation}
	Again by Schur complement and with Assumption 1, we obtain
	\begin{equation}
		\begin{aligned}
			\begin{bmatrix}
				-\frac{1}{\mu}\mathbf{C}+\tau I_N+\frac{\alpha c_1^2}{\tau}P^2+(\frac{\beta c_2^2}{\tau}+\frac{c}{\nu})LL^\top& 0 \\
				0& -\frac{1}{\nu}P
			\end{bmatrix} \\
			+\left[*\right]^\top{\mu}\mathbf{A}^{-1}
			\begin{bmatrix}
				\left(\begin{bmatrix}
					P \\ L^\top \\ 0
				\end{bmatrix}-\frac{1}{\mu}\mathbf{B}\right) \quad-\begin{bmatrix}
					0 \\ 0 \\ P
				\end{bmatrix}
			\end{bmatrix}\prec 0.
		\end{aligned}
	\end{equation}
	By calculation and also with \eqref{definition of zeta and Q}, matrix at the left hand side can be written as 
	\begin{equation}\label{Middle T matrix}
		\left[
			\begin{array}{c|c}
				\mathbf{T}_{11}& 
				-\mu\begin{bmatrix}
					P \\ L^\top \\ 0
				\end{bmatrix}^\top\mathbf{A}^{-1}\begin{bmatrix}
				0 \\ 0 \\ P
				\end{bmatrix}+\mathbf{B}^\top\mathbf{A}^{-1}\begin{bmatrix}
					0 \\ 0 \\ P
				\end{bmatrix} \\
				\hline
				*& 
				-\frac{1}{\nu}P+\mu\begin{bmatrix}
					0 \\ 0 \\ P
				\end{bmatrix}\mathbf{A}^{-1}\begin{bmatrix}
					0 \\ 0 \\ P
				\end{bmatrix}
			\end{array}
		\right]
	\end{equation}
	where
	\begin{equation}
		\begin{aligned}
			\mathbf{T}_{11}=\tau I_N+\frac{\alpha c_1^2}{\tau}P^2 +\left(\frac{\beta c_2^2}{\tau}+\frac{c}{\nu}\right)LL^\top+\frac{1}{\mu}\mathbf{Q}\\
			+\mu\begin{bmatrix}
				P \\ L^\top \\ 0
			\end{bmatrix}^\top\mathbf{A}^{-1}\begin{bmatrix}
				P \\ L^\top \\ 0
			\end{bmatrix}+
			\begin{bmatrix}
				P \\ L^\top \\ 0
			\end{bmatrix}^\top \boldsymbol{\zeta}+\boldsymbol{\zeta}^\top
			\begin{bmatrix}
				P \\ L^\top \\ 0
			\end{bmatrix}
		\end{aligned}.
	\end{equation}
	We can decompose \eqref{Middle T matrix} to different parts, thus obtain
	\begin{equation}\label{Petersen mu condition}
		\begin{aligned}
			&\begin{bmatrix}
				\boldsymbol{\zeta}^\top\begin{bmatrix}
					P \\ L^\top \\ 0
				\end{bmatrix}+\begin{bmatrix}
					P \\ L^\top \\ 0
				\end{bmatrix}^\top\boldsymbol{\zeta}& \quad -\boldsymbol{\zeta}^\top\begin{bmatrix}
					0 \\ 0 \\ P
				\end{bmatrix} \\
				*& 0
			\end{bmatrix} +\begin{bmatrix}
				\mathbf{R}& 0\\0& -\frac{1}{\nu}P
			\end{bmatrix} \\
			+&\mu\begin{bmatrix}
				\begin{bmatrix}
					P & L & 0
				\end{bmatrix}\mathbf{A}^{-\frac{1}{2}} \\
				-\begin{bmatrix}
					0 & 0 & P
				\end{bmatrix}\mathbf{A}^{-\frac{1}{2}}
			\end{bmatrix}\left[*\right]^\top+\frac{1}{\mu}
			\begin{bmatrix}
				\mathbf{Q}^{\frac{1}{2}}\\ 0
			\end{bmatrix}\left[*\right]^\top\prec 0,
		\end{aligned}
	\end{equation}
	where we define
	\begin{equation}
		\mathbf{R}=\tau I_N+\frac{\alpha c_1^2}{\tau}P^2 +\left(\frac{\beta c_2^2}{\tau}+\frac{c}{\nu}\right)LL^\top.
	\end{equation}
	Then by Lemma 2 (strict Petersen's lemma), we conclude that for any $\boldsymbol{\gamma}$ satisfying $ \|\boldsymbol{\gamma}\|\leq 1$ in the sense of \eqref{C=E reformulation}, the following inequality holds.
	\begin{equation}
		\begin{aligned}
			&\begin{bmatrix}
				\boldsymbol{\zeta}^\top\begin{bmatrix}
					P \\ L^\top \\ 0
				\end{bmatrix}+\begin{bmatrix}
					P \\ L^\top \\ 0
				\end{bmatrix}^\top\boldsymbol{\zeta}& \quad -\boldsymbol{\zeta}^\top\begin{bmatrix}
					0 \\ 0 \\ P
				\end{bmatrix} \\
				*& 0
			\end{bmatrix}+\begin{bmatrix}
				\mathbf{R}& 0\\0& -\frac{1}{\nu}P 
			\end{bmatrix}\\
			+&\begin{bmatrix}
			\mathbf{Q}^{\frac{1}{2}}\boldsymbol{\gamma}^\top\mathbf{A}^{-\frac{1}{2}}\begin{bmatrix}
				P \\ L^\top \\ 0
			\end{bmatrix}& 
			-\mathbf{Q}^{\frac{1}{2}}\boldsymbol{\gamma}^\top\mathbf{A}^{-\frac{1}{2}}
			\begin{bmatrix}
				0 \\ 0 \\ P
			\end{bmatrix} \\
			0& 0
			\end{bmatrix}+\left[*\right]^\top \prec 0 
		\end{aligned}
	\end{equation}
	
	Since $\mathcal{C}=\mathcal{E}$, $\mathbf{Z}=\boldsymbol{\zeta}+\mathbf{A}^{-\frac{1}{2}} \boldsymbol{\gamma}\mathbf{Q}^{\frac{1}{2}}$, we obtain
	\begin{equation}\label{Direct inequality with Z}
		\begin{bmatrix}
			\mathbf{R}+\mathbf{Z}^\top\begin{bmatrix}
				P \\ L^\top \\ 0
			\end{bmatrix}+\begin{bmatrix}
				P \\ L^\top \\ 0
			\end{bmatrix}^\top \mathbf{Z}& \quad-\mathbf{Z}^\top\begin{bmatrix}
				0 \\ 0 \\ P
			\end{bmatrix}\\
			*& -\frac{1}{\nu}P
		\end{bmatrix}\prec 0
	\end{equation}
	As $\mathbf{Z}^\top=[A \ B_0 \ B_1]$, the above inequality becomes
	\begin{equation}
		\begin{bmatrix}
			\mathbf{R}+PA^\top+AP+LB_0^\top+B_0L^\top& -B_1P\\
			-PB_1^\top& -\frac{1}{\nu}P
		\end{bmatrix}\prec 0.
	\end{equation}
	Finally by Schur complement, we conclude that for any $\mathbf{Z}=[A \ B_0 \ B_1]^\top\in \mathcal{C}$, the controller $u=L^\top P^{-1}z$ ensures the closed-loop stability, since 
	\begin{equation}\label{NMI m=1}
		\begin{aligned}
			PA^\top+AP+LB_0^\top+B_0L^\top+\nu B_1PB_1^\top+\tau I_N \\
			+\nu^{-1}cLL^\top+\tau^{-1}\left(\alpha c_1^2P^2+\beta c_2^2 LL^\top\right)\prec 0
		\end{aligned}
	\end{equation}
	always holds. The proof is completed. \qed
\end{pf}
\begin{remark}
	The negative definiteness of \eqref{Direct LMI} implies a semi-definite program (SDP) involving matrices $P,L$ and scalar parameters $\mu,\alpha,\beta,\tau,\nu$. To reduce the number of free parameters, certain scalars (e.g., $\alpha=\beta=2, c=1$) may be fixed a priori. The proof of Theorem 1 relies on two key tools: the Schur complement, offering a necessary and sufficient condition for determining the positive (negative) definiteness of a matrix, and Petersen's lemma, offering a necessary and sufficient condition for establishing the negative definiteness of a class of matrix inequalities. Notably, the conservatism of Theorem 1 stems entirely from Proposition 1's reliance on quadratic Lyapunov functions for robust stabilization of the lifted bilinear system \eqref{bilinear system no disturbance}, as the inequality \eqref{NMI m=1} inherently requires negative definiteness of the Lyapunov derivative. Future research could explore alternative stability analysis frameworks to mitigate conservatism in closed-loop guarantees.
\end{remark}


\section{Indirect Data-Driven Control Approach}
This section develops the indirect counterpart of the robust stabilization framework. Leveraging the most widely validated methodology for data-driven system identification based on the Koopman operator, we ensure the generality of the proposed theoretical results in practical applications.
\subsection{Data-Driven Identification via EDMD Algorithm}
EDMD algorithm has been widely utilized to approximate the Koopman operator and its generator, also leading to the lifted bilinear system. The convergence of EDMD algorithm has been proved in \cite{2017On}, with applications on a variety of control problems including robots \cite{SoftRobots2021TRO}, quadrotors \cite{Quadrotor2024Access}, smart grids \cite{Modularized2024IEEESmartGrid}, etc.

Despite the existence of its variants, fundamental theoretical principles and the most general methodology remain the same \cite{Strasser2024Koopman}. The collected data and notations are the same as \eqref{data set}, which lead to \eqref{Relations ZUV} and equivalently
\begin{equation}\label{identified relations}
	Z_1=\left[A\ B_0\ B_1\ \cdots \ B_m\right]W_0+D_0.
\end{equation}
Discarding the process disturbance, EDMD algorithm leads to the least square problem
\begin{equation}
    \min_{A,B_0,B_1,\cdots,B_m}\left\|Z_1-\left[A\ B_0\ B_1\ \cdots\ B_m\right]W_0\right\|
\end{equation}
which is solved by
\begin{equation}\label{identified z}
	\left[\hat{A}\ \hat{B}_0\ \hat{B}_1\ \cdots\ \hat{B}_m\right]:=\hat{\mathbf{Z}}^\top =Z_1W_0^\dagger
\end{equation}
and identifies the bilinear system \eqref{bilinear system no disturbance} as
\begin{equation}\label{identified z system}
	\frac{\mathrm{d}z}{\mathrm{d}t}=\hat{A}z+\hat{B}_0u+\sum_{i=1}^m u_i\hat{B}_iz +r(z,u).
\end{equation}
The identified matrix $\hat{A}$ is usually used to compute the Koopman eigenfunctions \cite{2019Data}. Assume that $\hat{A}$ is diagonalizable, i.e., $\exists V\in\mathbb{R}^{N\times N},\text{s.t.}\hat{\Lambda}=V^{-1}\hat{A}V$, where the diagonal elements of $\hat{\Lambda}$ are corresponding eigenvalues and the columns of $V=\left[ v_1 \ \cdots\ v_N\right]$ are also eigenvectors of $\hat{A}$. Then the approximated Koopman eigenfunctions are $\hat{\Phi}(x)=V^\top\Psi(x)=\left[ v_1^\top\Psi(x)\ \cdots\ v_N^\top\Psi(x) \right]$. In the case of non-diagonalizable $\hat{A}$, $\hat{\Lambda}$ is of Jordan canonical form, $v_j^\top\Psi(x), j=1,\cdots,N$ are generalized Koopman eigenfunctions \cite{TAC2025KoopmanHJB}. The corresponding identified dynamics
\begin{equation}
	\begin{aligned}\label{identified phi system}
		\frac{\mathrm{d}\hat{\Phi}(x)}{\mathrm{d}t}&=\hat{\Lambda}^\top\hat{\Phi}(x)+V^\top{B}_0u \\ &+\sum_{i=1}^m u_iV^\top\hat{B}_i(V^\top)^{-1}\hat{\Phi}(x) +\hat{r}\left(\hat{\Phi}(x),u\right)
	\end{aligned}
\end{equation}
is equivalent to \eqref{identified z system}.

Equations \eqref{identified relations}-\eqref{identified z} exhibit the general system identification methodology via EDMD algorithm. Note that $W_0$ is assumed to be of full row rank by Assumption~2. In fact, this assumption is standard and related to the notion of persistent excitation \cite{BISOFFI2022Petersen}, posing requirements on input signals $U_0$. For example, some existing works \cite{Strasser2024Koopman},\cite{2019Data} use zero and $m$ unit vectors $\{e_i\}_{i=0}^m$ as reference inputs. This method is a special case for the general methodology \eqref{identified relations}-\eqref{identified z}, since
\begin{equation}\label{data collection special case}
	W_0=\begin{bmatrix}
		Z_0 \\ U_0 \\ V_0^1 \\ V_0^2 \\ \vdots \\ V_0^m
	\end{bmatrix}=\begin{bmatrix}
		Z_{T_0}& Z_{T_1}& Z_{T_2}& \cdots& Z_{T_m} \\
		0& \mathbf{1}_{T_1}\otimes e_1& \mathbf{1}_{T_2}\otimes e_2& \cdots& \mathbf{1}_{T_m}\otimes e_m \\
		0& Z_{T_1}& 0& \cdots& 0 \\ 0& 0& Z_{T_2}& \cdots& 0 \\
		\vdots& \vdots& \vdots& \ddots& \vdots \\ 0& 0& 0& 0& Z_{T_m} 
	\end{bmatrix}
\end{equation}
where $Z_{T_j}$ denotes data with $T_j$ samples, $\mathbf{1}_{T_j}$ is a $T_j$-dimensional row vector with all $1$. Clearly, if $Z_{T_j},j=0,1,2,\cdots,m$ are all of full row rank with sufficient samples $T_j$, $W_0$ is of full row rank, satisfying Assumption 2.

\subsection{Main Results for Indirect Data-Driven Control}
Now we introduce our main result for indirect data-driven control approach.
\begin{theorem}
	Consider system \eqref{identified z system} identified with data sets $Z_0,U_0,V_0^1$ collected from disturbed system \eqref{bilinear system z disturbance} (or equivalently \eqref{original system}) satisfying Assumption 2. The lifted-state feedback controller $u=L^\top P^{-1}z$ ensures closed-loop stability for \eqref{bilinear system z disturbance} also \eqref{original system} with respect to any possible disturbance, if there exist $0\prec P=P^\top\in\mathbb{R}^{N\times N}, L\in\mathbb{R}^{N\times 1}$, and positive scalars $\mu,\alpha,\beta,\tau,\nu$ where $\frac{1}{\alpha}+\frac{1}{\beta}=1$, such that $P\prec\frac{\rho}{\mu\nu}I_N$ and 
	\begin{equation}\label{InDirect NMI}
		\begin{aligned}
			&P\hat{A}^\top+\hat{A}P+L\hat{B}_0^\top+\hat{B}_0L^\top+\mathbf{R}+\mu^{-1}\Delta\Delta^\top\\
			+&\frac{\mu}{\rho}(P^2+LL^\top)+\nu \hat{B}_1P\left(I_N-\frac{\mu\nu}{\rho}P\right)^{-1}\hat{B}_1^\top  \prec 0
		\end{aligned}
	\end{equation}
	where $\rho=1/\lambda_{\min}(W_0W_0^\top)$. Moreover, the region of attraction contains the set $\mathbb{X}_{RoA}=\{x\in \mathbb{X}:z^\top P^{-1}z=\Psi(x)^\top P^{-1}\Psi(x)\leq c\}$.
\end{theorem}
\begin{pf}
	We start with \eqref{Direct inequality with Z}, which is equivalent to \eqref{NMI} if $m=1$. The system is identified by \eqref{identified z} via EDMD. But taking the process disturbance into account, the exact solution is
	\begin{equation}\label{Zexact}
		\left[A \ B_0 \ B_1 \ \cdots \ B_m \right]=\mathbf{Z}^\top=(Z_1-D_0)W_0^\dagger =\hat{\mathbf{Z}}^\top -D_0W_0^\dagger.
	\end{equation}
	Without any knowledge of disturbance, we can only use the least-square solution \eqref{identified z} for nonlinear stabilization. However, we should actually use \eqref{Zexact} to stabilize the original nonlinear system. Therefore, if \eqref{Direct inequality with Z} always holds for any disturbance sequence $D_0\in\mathcal{D}$ and corresponding $\mathbf{Z}^\top=\hat{\mathbf{Z}}^\top -D_0W_0^\dagger$, we can conclude that the controller actually stabilizes \eqref{original system}.
	
	By the Schur complement, \eqref{InDirect NMI} is equivalent to
	\begin{equation}\label{three terms of indirect}
		\begin{aligned}
			&\left[
			\begin{array}{c|c}
				\begin{matrix}
					P\hat{A}^\top+\hat{A}P+L\hat{B}_0^\top+\hat{B}_0L^\top \\
					+\mathbf{R}+\frac{\mu}{\rho}(P^2+LL^\top)+\mu^{-1}\Delta\Delta^\top
				\end{matrix}& -B_1P\\
				\hline
				-PB_1^\top& -\frac{1}{\nu}P+\frac{\mu}{\rho} P^2\\
			\end{array}
			\right] \\
			=&\begin{bmatrix}
				\mathbf{R}+\hat{\mathbf{Z}}^\top\begin{bmatrix}
					P \\ L^\top \\ 0
				\end{bmatrix}+\begin{bmatrix}
					P \\ L^\top \\ 0
				\end{bmatrix}^\top \hat{\mathbf{Z}}& \quad -\hat{\mathbf{Z}}^\top\begin{bmatrix}
					0 \\ 0 \\ P
				\end{bmatrix}\\
				*& -\frac{1}{\nu}P
			\end{bmatrix} \\
			+&\frac{\mu}{\rho}\begin{bmatrix}
				P^2+LL^\top& 0\\ 0& P^2
			\end{bmatrix}+\frac{1}{\mu}
			\begin{bmatrix}
				\Delta\Delta^\top& 0\\0& 0
			\end{bmatrix}
			\prec 0.
		\end{aligned}
	\end{equation}
	For the second term, we have
	\begin{equation}\label{second term of indirect}
		\begin{aligned}
			&\frac{\mu}{\rho}\begin{bmatrix}
				P^2+LL^\top& 0\\ 0& P^2
			\end{bmatrix}=\frac{\mu}{\rho}
			\begin{bmatrix}
				\begin{bmatrix}
					P \ L \ 0
				\end{bmatrix}\\
				-\begin{bmatrix}
					0 \ 0 \ P
				\end{bmatrix}
			\end{bmatrix}\left[*\right]^\top
			\\
			\succeq& \mu
			\begin{bmatrix}
				\begin{bmatrix}
					P \ L \ 0
				\end{bmatrix}\\
				-\begin{bmatrix}
					0 \ 0 \ P
				\end{bmatrix}
			\end{bmatrix}\left(W_0W_0^\top\right)^{-1}\left[*\right]^\top \\
			=&\mu\begin{bmatrix}
				\begin{bmatrix}
					P \\ L^\top \\ 0
				\end{bmatrix}^\top\mathbf{A}^{-1}\begin{bmatrix}
					P \\ L^\top \\ 0
				\end{bmatrix}& \quad*\\
				-\begin{bmatrix}
					0 \\ 0 \\ P
				\end{bmatrix}^\top\mathbf{A}^{-1}\begin{bmatrix}
					P \\ L^\top \\ 0
				\end{bmatrix}& \quad\begin{bmatrix}
					0 \\ 0 \\ P
				\end{bmatrix}^\top\mathbf{A}^{-1}\begin{bmatrix}
					0 \\ 0 \\ P
				\end{bmatrix}
			\end{bmatrix}
		\end{aligned}
	\end{equation}
	since $\rho=\lambda_{\max}((W_0W_0^\top)^{-1})=\frac{1}{\lambda_{\min}(W_0W_0^\top)}$.
	
    Combining \eqref{second term of indirect} with \eqref{three terms of indirect}, we obtain
    \begin{equation}
        \begin{aligned}
            \begin{bmatrix}
			\mathbf{R}+\hat{\mathbf{Z}}^\top\begin{bmatrix}
				P \\ L^\top \\ 0
			\end{bmatrix}+\begin{bmatrix}
				P \\ L^\top \\ 0
			\end{bmatrix}^\top \hat{\mathbf{Z}}& \quad-\hat{\mathbf{Z}}^\top\begin{bmatrix}
				0 \\ 0 \\ P
			\end{bmatrix}\\
				*& -\frac{1}{\nu}P
		\end{bmatrix} \\ +\mu
			\begin{bmatrix}
				\begin{bmatrix}
					P \ L \ 0
				\end{bmatrix}\\
				-\begin{bmatrix}
					0 \ 0 \ P
				\end{bmatrix}
			\end{bmatrix}\mathbf{A}^{-1}\left[*\right]^\top+\frac{1}{\mu}
			\begin{bmatrix}
				\Delta\Delta^\top& 0\\0& 0
			\end{bmatrix}
			\prec 0.
        \end{aligned}
    \end{equation}
    Using Lemma 2, we conclude that for any $D_0\in\mathcal{D}$, 
	\begin{equation}\label{indirect Petersen}
		\begin{aligned}
			&\begin{bmatrix}
				\mathbf{R}+\hat{\mathbf{Z}}^\top\begin{bmatrix}
					P \\ L^\top \\ 0
				\end{bmatrix}+\begin{bmatrix}
					P \\ L^\top \\ 0
				\end{bmatrix}^\top \hat{\mathbf{Z}}& \quad -\hat{\mathbf{Z}}^\top\begin{bmatrix}
					0 \\ 0 \\ P
				\end{bmatrix}\\
				*& -\frac{1}{\nu}P
			\end{bmatrix}\\
			+&\begin{bmatrix}
				-\begin{bmatrix}
					P & L & 0
				\end{bmatrix}(W_0^\top)^\dagger& \quad 0\\ \begin{bmatrix}
					0 & 0 & P
				\end{bmatrix}(W_0^\top)^\dagger& \quad 0
			\end{bmatrix}\begin{bmatrix}
				D_0^\top \\ 0
			\end{bmatrix} \\
			+&\left[D_0\ 0\right]
			\begin{bmatrix}
				-W_0^\dagger\begin{bmatrix}
					P \\ L^\top \\ 0
				\end{bmatrix}& \quad W_0^\dagger\begin{bmatrix}
					0 \\ 0 \\ P
				\end{bmatrix} \\ 0& \quad 0
			\end{bmatrix}\prec 0
		\end{aligned}
	\end{equation}
	holds since the row full rank matrix $W_0$ satisfies $W_0^\dagger=W_0^\top(W_0W_0^\top)^{-1}$, the column full rank matrix $(W_0^\top)^\dagger=(W_0W_0^\top)^{-1}W_0$, and $(W_0^\top)^\dagger W_0^\dagger=\mathbf{A}^{-1}$. Now with \eqref{Zexact}, we obtain \eqref{Direct inequality with Z} under any $D_0\in\mathcal{D}$ and corresponding any possible exact solution $\mathbf{Z}=\hat{\mathbf{Z}}-(W_0^\top)^\dagger D_0^\top$. The proof is completed. \qed
\end{pf}
\begin{remark}
	A critical analysis of Theorem 2’s proof reveals that all derivation steps, with the exception of the conservatism introduced by the minimum eigenvalue of $\mathbf{A}=W_0W_0^\top$ in \eqref{second term of indirect}, proceed via equivalent transformations. Consequently, akin to Theorem 1, the dominant source of conservatism persists in the limitations of Proposition 1’s reliance on quadratic Lyapunov functions, leaving room for future refinements through alternative strategies. A critical observation lies in the coherence between the perturbed and nominal control designs: the disturbance-robust matrix inequality \eqref{InDirect NMI} inherently generalizes its disturbance-free counterpart through a parametric extension. Specifically, under disturbance-free conditions (i.e., $\Delta\Delta^\top=0$), setting $\mu=0$ recovers \eqref{NMI} from Proposition 1, demonstrating backward compatibility with the nominal stability framework. Meanwhile, \eqref{InDirect NMI} can also be transformed to an LMI
	\begin{equation}\label{Indirect LMI}
		\left[
		\begin{array}{c|c}
			\begin{matrix}
				\tilde{E}_{11}& 0& -B_1P& -P& -L \\
				0& -\frac{\rho}{\mu}I_N& P& 0& 0 \\ -PB_1^\top& P& -\frac{1}{\nu}P& 0& 0\\
				-P& 0& 0& -\frac{\rho}{\mu}I_N& 0\\ -L^\top& 0& 0& 0& -\frac{\rho}{\mu}
			\end{matrix}& \begin{matrix}
				L& P& L\\0& 0& 0\\0& 0& 0\\0& 0& 0\\0& 0& 0
			\end{matrix}
			\\ \hline
			*& E_{22}			
		\end{array}
		\right]\prec 0
	\end{equation}
	where $\tilde{E}_{11}=P\hat{A}^\top+\hat{A}P+L\hat{B}_0^\top+\hat{B}_0L^\top +\tau I_N+\mu^{-1}\Delta\Delta^\top$ and $E_{22}$ is defined the same as \eqref{Definition E12E22}.
\end{remark}

\section{Discussions on the Proposed Results}
Sections 3 and 4 investigate the direct and indirect data-driven control approaches respectively, with main results presented in Theorems 1 and 2. This section provides a concise yet critical discussion of these findings. First, we rationalize the investigation of controllers with linear dependence on the lifted state within the robust stabilization framework. Moreover, we highlight novel insights and conceptual innovations emerging from comparative analysis with existing literature.

\subsection{Validity of the Lifted-State Feedback Controller}
In Theorems 1 and 2, designed controllers given by $u=Kz=L^\top P^{-1}z$ exhibit a linear dependence on the lifted-state. A natural question arises as to whether this structure is sufficient for addressing bilinear system control problems, as exemplified by the infinite-horizon optimal control \cite{1995Linear}. However, investigating the robust stabilization problem using the designed controllers $u=Kz$ remains to be justified. The following theorem seeks to confirm that the linear component with respect to $z=\Psi(x)$ is far more critical to guarantee closed-loop stability.
\begin{theorem}
	Consider the lifted bilinear system \eqref{bilinear system no disturbance}. Assume the feedback controller that linearly dependent on the lifted-state $u=Kz$ obtained by \eqref{NMI} stabilizes \eqref{bilinear system no disturbance}. Then if an additional nonlinear term $u_n(z)$ is added to the original controller resulting in a modified one $u=Kz+u_n(z)$, where $u_n(z)$ is a high-order small term with respect to $z$, i.e.,
	\begin{equation}\label{high order small}
		\lim_{z\rightarrow0}\frac{\|u_n(z)\|}{\|z\|}=0,
	\end{equation}
	the closed-loop system controlled by $u=Kz+u_n(z)$ is also locally stable.
\end{theorem}
\begin{pf}
	Denote $u_l(z)=Kz$ as the linear part of $z$ in the modified controller $u=Kz+u_n(z) =u_l(z)+u_n(z)$ where $K=L^\top P^{-1}$ is obtained by Proposition 1. With its proof in Appendix A, the solution of \eqref{NMI} leads to a Lyapunov function $V(z)=z^\top P^{-1}z$ satisfying $\frac{\mathrm{d}V}{\mathrm{d}t}<0$ within $\mathbb{X}_{RoA}$. After adding the nonlinear term $u_n(z)$, the time derivative of $V(z)$ becomes
	\begin{equation}\label{dVdt}
		\begin{aligned}
			\frac{\mathrm{d}V}{\mathrm{d}t}&=z^\top P^{-1}\left(Az+B_0u_l(z) +\sum_{i=1}^mu_{l_i}(z)B_iz\right) \\
			&+\left(Az+B_0u_l(z)+\sum_{i=1}^mu_{l_i}(z)B_iz\right)^\top P^{-1}z \\
			&+2z^\top P^{-1}\left(B_0u_n(z)+\sum_{i=1}^mu_{n_i}(z)B_iz+r(z,u)\right)
		\end{aligned}
	\end{equation}
	where $u_{l_i}(z),u_{n_i}(z), i=1,\cdots,m$ denote the $i$-th component of $u_l(z),u_n(z)$ respectively, and
	\begin{equation}
		\|r(z,u)\|\leq c_1\|z\|+c_2\|u_l(z)\|+c_2\|u_n(z)\|.
	\end{equation}
	Now decompose it into two parts, $r(z,u)=r_l(z,u)+r_n(z,u)=D_l(z,u)r(z,u)+D_n(z,u)r(z,u)$ where
	\begin{equation}
		\begin{aligned}
			D_l(z,u)=\frac{c_1\|z\|+c_2\|u_l\|}{c_1\|z\|+c_2\|u_l\|+c_2\|u_n\|},\\
			D_n(z,u)=\frac{c_2\|u_n\|}{c_1\|z\|+c_2\|u_l\|+c_2\|u_n\|}.
		\end{aligned}
	\end{equation}
	Additionally, we define $D_l(0,0)=1$ and $D_n(0,0)=0$ to enforce the limit-based continuity at $z=u=0$. As $u_l$ is linear with respect to $z$, and $u_n=u-u_l$, both $D_l$ and $D_n$ are functions of $z$ and $u$, apparently
	\begin{subequations}\label{erroreq:main}
		\begin{align}
			&\|r_l(z,u)\|\leq c_1\|z\|+c_2\|u_l(z)\|,\label{erroreq:linear} \\ &\|r_n(z,u)\|\leq c_2\|u_n(z)\|.\label{erroreq:nonlinear}
		\end{align}
	\end{subequations}
	As proved in Proposition 1, the matrix inequality \eqref{NMI} for stabilization and corresponding controller $u_l(z)=Kz$ allow the error term to satisfy any possible form as long as it is bounded by \eqref{erroreq:linear}. Therefore, \eqref{dVdt} can be written as
	\begin{equation}\label{dVdt linear and nonlinear}
		\begin{aligned}
			\frac{\mathrm{d}V}{\mathrm{d}t}=\frac{\mathrm{d}V_0}{\mathrm{d}t}+2z^\top P^{-1}\Big(B_0u_n(z) \\ +\sum_{i=1}^mu_{n_i}(z)B_iz+r_n(z,u)\Big)
		\end{aligned}
	\end{equation}
	where $\frac{\mathrm{d}V_0}{\mathrm{d}t}=z^\top P^{-1}(Az+B_0Kz +\sum_{i=1}^mu_{l_i}(z)B_iz+r_l(z,u))+(Az+B_0Kz+\sum_{i=1}^mu_{l_i}(z)B_iz+r_l(z,u))^\top P^{-1}z$ corresponds to the linear component $u_l(z)=Kz$. Furthermore, since Proposition~1 achieves exponential stability \cite{khalil1996nonlinear} within the region of attraction $\mathbb{X}_{RoA}$, there exist $\eta>0$ and $\delta_1>0$, such that
    \begin{equation}
        \frac{\mathrm{d}V_0}{\mathrm{d}t}\leq-\eta z^\top z,\ \forall z\in\mathbf{B}_{\delta_1} =\{z\in\mathbb{R}^N:z^\top z\leq\delta_1\}.
    \end{equation}
    The remainder term of $\frac{\mathrm{d}V}{\mathrm{d}t}$, with \eqref{high order small}, is actually a higher-order small quantity, then for any $\varepsilon>0$, there exists $\delta_2>0$ such that $\forall z\in\mathbf{B}_{\delta_2}=\{z\in\mathbb{R}^N:z^\top z\leq\delta_2\}$,
	\begin{equation}
		\left\|B_0u_n(z) \\ +\sum_{i=1}^mu_{n_i}(z)B_iz+r_n(z,u)\right\|\leq \varepsilon \|z\|.
	\end{equation}
    We order that $\varepsilon=\frac{\eta}{3\lambda_{max}(P^{-1})}$. With \eqref{dVdt linear and nonlinear}, we claim that
    \begin{equation}
        \frac{\mathrm{d}V}{\mathrm{d}t}\leq-\frac{\eta}{3}z^\top z,\ \forall z\in \mathbf{B}_{\delta_1}\cap \mathbf{B}_{\delta_2}.
    \end{equation}
    In other words, $dV/dt<0$ holds at least in a small region sufficiently close to the origin $z=0$ where also $x=0$. The proof is completed. \qed
\end{pf}
Theorem 3 validates the use of linear lifted-state controllers $u=Kz$ for stabilizing both the lifted bilinear system and original nonlinear counterpart, which also underscores the analytical role of our main results mainly composed of Theorems 1 and 2. In practice, the controller can sometimes be decomposed into linear and nonlinear components with respect to $z$ under appropriate problem formulations. For instance, procedure one in \cite{TAC2025KoopmanHJB} utilizes the Koopman eigenfunctions $\Phi(x)$ to approximate the solution of HJB equation $V(x)=\frac{1}{2} \Phi(x)^\top L\Phi(x)$ where $L$ is obtained from the Riccati equation in \cite[Proposition 5]{TAC2025KoopmanHJB}, leading to the optimal controller
\begin{equation}
    u^*=-R^{-1}g^\top(x) \left(\frac{\partial\Phi}{\partial x}\right)^\top L\Phi(x).
\end{equation}
where $g(x)=[g_1(x)\ g_2(x)\ \cdots \ g_m(x)]^\top$. To obtain the linear component with respect to $\Phi(x)$, one need only extract the constant term in $g(x)$ and $\frac{\partial\Phi}{\partial x}$. As established in Theorem 3, the linear component $u_l$ is of greater relevance to closed-loop stability, whose stabilizability can be analyzed with Theorems 1 and 2. The nonlinear component $u_n$ is likely to influence the region of attraction, which is an issue that merits further investigation in future work, e.g., using linear parameter varying (LPV) method \cite{item_462da11ab7204e0c877cf583ed5f5d74} or extending the results to other controller design reducing conservatism \cite{STRASSER2025101286}. Furthermore, it is worth noting that the general optimal control framework leveraging the Koopman operator represents a promising direction for future research, for example, considering effects of uncertainties on optimal control problems. 

\subsection{Comparison with Existing Works}
Centering on the impacts of uncertainties on data-driven nonlinear stabilization leveraging the Koopman operator, Theorems 1 and 2 are presented herein. Both direct and indirect data-driven control methodologies are examined. Compared with existing results, several conceptual innovations and insights are synthesized as follows.

\textbf{Preserving the generality is of paramount importance.} As detailed in Section 3.2, different methodologies are integrated into a unified framework, addressing approximation error and process disturbance based on their intrinsic properties. If needed, the approximation error can also be handled via the process disturbance approach, for example, \cite{CDC2022You} defines a residual dataset with a bounded energy assumption
\begin{equation}
    R_0=[r_0\ r_1\ \cdots\ r_{T-1}],\ R_0^\top R_0<\overline{R}
\end{equation}
which is similar to the method in \eqref{definition D0} and \eqref{Assumption 1}. Furthermore, the state feedback control follows a more general structure $u=L^\top P^{-1}z$ compared with a relatively special form $u=U_0H(Z_0H)^{-1}z$, since $H$ is solved by an LMI under some additional assumptions and matrices $L,P$ admit more flexible tuning of controller design. Notably, no strict assumption is imposed on the dictionary functions \(\Psi(x)\) or the lifted bilinear system \eqref{bilinear system z disturbance}, collectively reducing limitations of theoretical framework in practice. 

\textbf{Mitigating conservatism via targeted efforts remains a critical pursuit.} To this end, we depart from reasoning in references like \cite{BISOFFI2020Bilinear}, which studies data-driven stabilization of bilinear systems under \(\|B_1\|\leq\delta_1\). Applying the strict Petersen's lemma (Lemma 2 in this paper), matrix inequality of the form (see \cite[Lemma 4]{BISOFFI2020Bilinear})
\begin{equation}\label{nonequivalence:1}
	\mathbf{C}+\mathbf{M}\frac{B_1}{\delta_1}\mathbf{N}^\top+\mathbf{N}\frac{B_1^\top}{\delta_1}\mathbf{M}^\top\prec 0
\end{equation}
leads to the existence of \(\mu>0\) such that 
\begin{equation}\label{nonequivalence:2}
	\mathbf{C}+\mu\mathbf{M}\mathbf{M}^\top+\mu^{-1}\mathbf{N}\mathbf{N}^\top\prec 0.
\end{equation}
This introduces non-equivalence: while \eqref{nonequivalence:2} holds for all \(B_1\) with \(\|B_1\|\leq\delta_1\), the identified \(B_i\) ( \(i=1,\cdots,m\)) are relatively fixed for a specific nonlinear system \eqref{original system}, making the matrix inequality \eqref{nonequivalence:2} more conservative than the required condition \eqref{nonequivalence:1}. In contrast, we apply Petersen's lemma to address process disturbances, enabling Theorems 1 and 2 to accommodate any disturbance form within the required energy bound.

\textbf{Conveying dual-functional values constitutes the core philosophy.} The proposed results serve as an analytical tool, as exemplified in Section 5.1, and naturally enable controller synthesis. To simultaneously ensure stabilizability and good performance, one may first use Theorem 1 or 2 to design the linear component (guaranteeing closed-loop stability), and then carefully tailor the nonlinear component for performance optimization. This two-stage approach, as evidenced by [18], has shown significant potential for further development.


\section{Numerical Examples}
In this section, we conduct the simulation with a physical example, an inverted pendulum. Consider the dynamics
\begin{equation}\label{numerical system}
	\begin{bmatrix}
		\dot{x}_1 \\ \dot{x}_2
	\end{bmatrix}=\begin{bmatrix}
		x_2 \\ -\frac{b}{ml^2}x_2+\frac{g}{l}\sin(x_1)
	\end{bmatrix}+\begin{bmatrix}
		0 \\ 1
	\end{bmatrix}u+d
\end{equation}
with mass $m=1$, length $l=1$, rotational friction coefficient $b=0.01$, and gravitational constant $g=9.81$. It is noteworthy that system dynamics is used only for data collection, while the controller design and analysis follow data-driven method. The process disturbance satisifies
\begin{equation}\label{disturbance numerical}
	d=\begin{bmatrix}
		\sqrt{\delta}\cos(2\pi\cdot0.6t) \\ \sqrt{\delta}\sin(2\pi\cdot0.6t)
	\end{bmatrix}
\end{equation}
with the bounded constant $\delta>0$ which satisifies Assumption 1 with $\Delta=\sqrt{T\delta}I$ in (21).

To systematically validate our theoretical findings, we implement cross-validation simulations that rigorously examines both direct and indirect data-driven control approaches. Specifically, the feedback controller synthesized through the direct approach undergoes closed-loop stability analysis using the indirect approach (as shown in Section 6.1), and vice versa (in Section 6.2). This mutual validation not only corroborates the consistency of two approaches, but also highlights dual-functional values of proposed theoretical results in robust controller design and closed-loop stability verification.

\subsection{Direct Data-Driven Control}
In this subsection, we aim to verify the effectiveness of direct data-driven control approach in Theorem 1. The dictionary of observables is selected as
\begin{equation}\label{numerical lifting}
	z=\Psi(x)=[x_1\ \ x_2\ \ \sin(x_1)\ \ x_2\cos(x_1)]^\top.
\end{equation}
We collect the data generated under zero and unit vectors, leading to the dataset specified in \eqref{data collection special case}. The data samples are uniformly sampled from the state space $\mathbb{X}=[-2,10]^2$ under process disturbance given in \eqref{disturbance numerical} with $\delta=0.01$, where the total number of data points is $T=1000$. Additionally, the matrix $W_0$ is verified to be of full rank. The approximation error $r(z,u)$ is bounded by \eqref{error bound} with $c_1=c_2=0.1$.

The negative definiteness condition for matrix \eqref{Direct LMI} translates to an LMI, which is solved using the YALMIP toolbox \cite{1393890} in MATLAB. To ensure the LMI is solvable with the widely used SDP solver MOSEK, we adopt a line search procedure \cite{BISOFFI2020Bilinear}, fixing the scalar variable $\nu$, solving the LMI, and iterating on the selection of $\nu$ if necessary. Accordingly, a feasible controller
\begin{equation}\label{direct controller}
	u_{\text{direct}}=[ -2.1061\ -3.4635\ -13.4264\ -3.0480]\Psi(x)
\end{equation}
is obtained with the positive scalars $\alpha=3,\ \beta=3/2,\ c=5,\ \nu=10,\ \tau=0.0267,\ \mu=46.22$. As shown in Fig. 1, closed-loop trajectories are demonstrated to be convergent. Through cross-validation with the indirect data-driven control approach, the controller \eqref{direct controller}, along with the corresponding matrices 
\begin{equation}
	\begin{aligned}
		P=\begin{bmatrix}
			0.0239 & -0.0168 & 0.0011 & -0.0021 \\
			-0.0168 &  0.0289 & -0.0027 & -0.0033 \\
			0.0011 & -0.0027 &  0.0012 & -0.0030 \\
			-0.0021 & -0.0033 & -0.0030 & 0.0353
		\end{bmatrix},\\ L=\begin{bmatrix}
			-0.0007& -0.0178& 0.0005& -0.0519
		\end{bmatrix}^\top
	\end{aligned}
\end{equation}
and scalars $\alpha,\beta,\mu,\tau,\nu$ also satisfies the matrix inequality \eqref{InDirect NMI} equivalently \eqref{Indirect LMI}, where $\hat{A},\hat{B}_0,\hat{B}_1$ are obtained from \eqref{identified z}. 

\begin{figure}[htbp]
    \centering
    \includegraphics[width=0.48\textwidth]{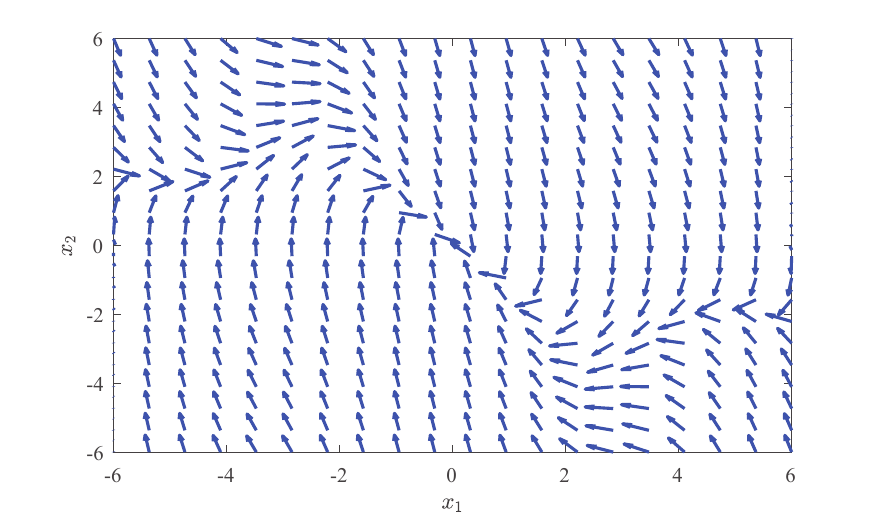}
    \caption{Phase plot of system \eqref{numerical system} under the direct data-driven control law \eqref{direct controller}, with arrows indicating the closed-loop vector field. Trajectories from distinct initial conditions converge asymptotically to the origin.}
    \label{fig:figure1}
\end{figure}

\subsection{Indirect Data-Driven Control}
To further validate the generality of proposed results, a different data collection method compared with Section 6.1 is employed. As illustrated in Fig. 2, data are generated from a randomly selected initial state under the control of a linear chirp signal with amplitude 2 and frequency ranging from 0 to 0.8. The evolution of system \eqref{numerical system} is subject to the process disturbance defined in \eqref{disturbance numerical} with $\delta=0.0005$. The dataset consists of $T=1000$ samples with a sampling interval $\Delta t=0.01$. It is verified that the resulting data matrix $W_0$ is of full rank. With EDMD algorithm \eqref{identified relations}-\eqref{identified z}, the lifted bilinear system \eqref{identified z system} is identified from the collected data where
\begin{equation}
	\begin{aligned}
		\hat{A}&=\begin{bmatrix}
			0& 1.00& 0& 0\\ 0& -0.01& 9.81& 0\\ 0& 0& 0& 1.00 \\ -0.20& 0.50& -9.03& 0.37
		\end{bmatrix},\hat{B}_0=\begin{bmatrix}
			-0.07\\ 1.02\\ 0.05\\ 18.99
		\end{bmatrix},\\ \hat{B}_1&=\begin{bmatrix}
			0.02& 0& 0.03& 0\\ 0& 0& 0& 0\\ -0.02& 0& -0.02& 0\\ -6.30& 0.39& -6.62& 0.16
		\end{bmatrix}.
	\end{aligned}
\end{equation}
\begin{figure}[htbp]
	\centering
	\includegraphics[width=0.50\textwidth]{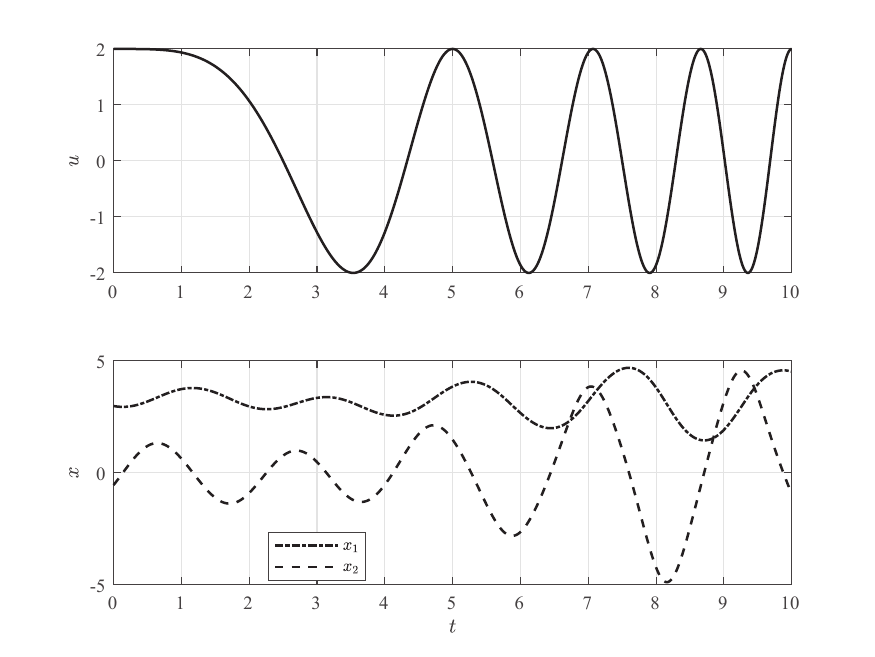}
	\caption{Input signal and resulting state trajectories used for system identification. A linear chirp signal is applied as control input. The state evolution is recorded from a randomly selected initial condition under process disturbance.}
	\label{fig:figure2}
\end{figure}

The approximation error $r(z,u)$ is bounded by \eqref{error bound} with $c_1=c_2=0.1$. Solving \eqref{Indirect LMI} follows similar procedures in Section 6.1, yielding a feasible controller
\begin{equation}\label{indirect controller}
	u_{\text{indirect}}=[-2.1405 \ -3.7326 \ -16.3943 \ -2.6147]\Psi(x)
\end{equation}
with the positive scalars $\alpha=2,\ \beta=2,\ c=5,\ \nu=11.11,\ \tau=0.0047,\ \mu=0.7699$. As shown in Fig. 3, closed-loop trajectories from multiple initial points converge to the origin. Implementing a cross validation with direct data-driven control approach, the controller \eqref{indirect controller} and corresponding matrices $P,L$ and also ensure the negative definiteness of \eqref{Direct LMI} and satisfy Theorem 1.
\begin{figure}[htbp]
	\centering
	\includegraphics[width=0.48\textwidth]{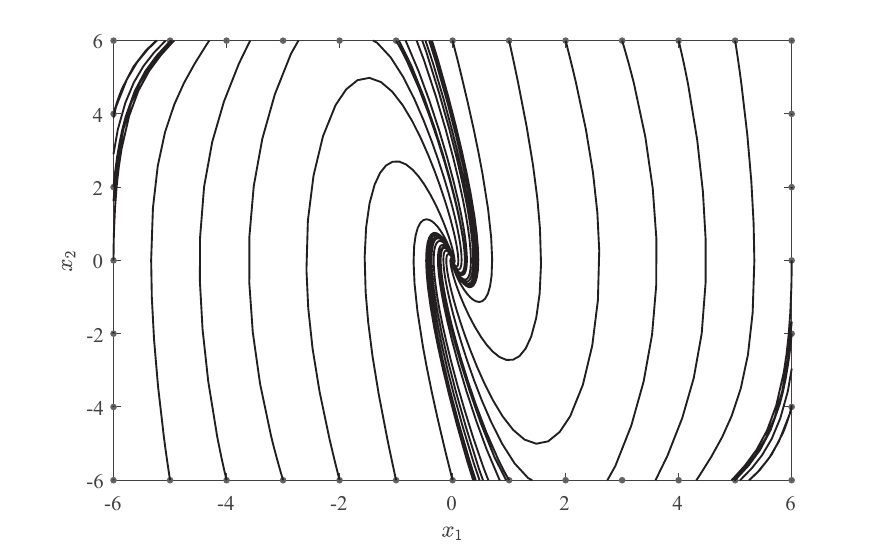}
	\caption{Closed-loop trajectories of \eqref{numerical system} under control law \eqref{indirect controller} where initial points, denoted by the solid circular markers, are uniformly selected along the boundaries of square domain $[-6,6]^2$. It can be observed that all trajectories converge to the origin.}
	\label{fig:figure3}
\end{figure}


\section{Conclusions}
We integrate three kinds of uncertainties into a unified framework and solve the robust stabilization problem in both direct and indirect data-driven control via the Koopman operator. Corresponding results are given in Theorems 1 and 2 respectively, which not only provide fundamental theoretical guarantees for robustness against composite uncertainties but also enable practical controller synthesis through LMIs. These advances yield significantly broader applicability and reduced conservatism compared to existing approaches.

Future works will focus on demonstrating these theoretical results in applications. Furthermore, providing more solid theoretical supports for data-driven control applications using the Koopman operator, especially in integrating with major domains like optimal control and adaptive control, is a research direction worth pursuing.
\begin{ack}                               
We would like to express our sincere gratitude to the National Natural Science Foundation of China (NSFC) for financial support under grants T2121002, U24A20266, and 62173006.  
\end{ack}

\bibliographystyle{plain}        
\bibliography{Automatica_References}           

\appendix
\section{Proof of Proposition 1}
The proof partially refers to the thoughts in \cite{Strasser2024Koopman}. Under the existence of $P,L$, construct a Lyapunov function $V(x)=z^\top P^{-1}z$. As mentioned $z(0)=0$ leads to $V(0)=0$, also the quadratic form promises $V(x)>0, x\neq 0$. Let $K=L^\top P^{-1}, A_K=A+B_0K$. By the well-known Schur complement lemma \cite{Herrmann2007}, inequality \eqref{NMI} is equivalent to the positive definiteness of
\begin{equation}
	\begin{bmatrix}
		-PA_K^\top-A_KP-\tau I_{N-1}&  -L&  -P&  -L&  B_1P \\
		-L^\top& \frac{\nu}{c}I_m&  0&  0&  0 \\
		-P& 0& \frac{\tau}{\alpha c_1^2}I_{N-1}& 0& 0 \\
		-L^\top& 0& 0& \frac{\tau}{\beta c_2^2}I_m& 0 \\
		PB_1^\top& 0& 0& 0& \frac{1}{\nu} P
	\end{bmatrix}.\label{LMI}
\end{equation}
Again by Schur complement, the above matrix becomes
\begin{equation*}
	\begin{aligned}
		\begin{bmatrix}
			-PA_K^\top-A_K P-\tau I_{N-1}& -L& -P& -L \\
			-L^\top& \frac{\nu}{c}I_m& 0& 0 \\
			-P& 0& \frac{\tau}{\alpha c_1^2}I_{N-1}& 0 \\
			-L^\top& 0& 0& \frac{\tau}{\beta c_2^2}I_m
		\end{bmatrix} \\ -\nu\begin{bmatrix}
			B_1 \\ 0 \\ 0 \\ 0
		\end{bmatrix}P\begin{bmatrix}
			B_1 \\ 0 \\ 0 \\ 0
		\end{bmatrix}^\top=
		\begin{bmatrix}
			-PA_K^\top-A_K P& -L& -P& -L \\
			-L^\top& 0& 0& 0 \\
			-P& 0& 0& 0 \\
			-L^\top& 0& 0& 0
		\end{bmatrix}
	\end{aligned}
\end{equation*}
\begin{equation*}
	+\begin{bmatrix}
		-\nu B_1PB_1^\top& 0& 0& 0 \\
		0& \frac{\nu}{c}I_m& 0& 0 \\
		0& 0& 0& 0 \\
		0& 0& 0& 0
	\end{bmatrix}
\end{equation*}
\begin{equation*}
	\begin{aligned}
		+\begin{bmatrix}
			-\tau I_{N-1}& 0& 0& 0 \\
			0& 0& 0& 0 \\
			0& 0& \frac{\tau}{\alpha c_1^2}I_{N-1}& 0 \\
			0& 0& 0& \frac{\tau}{\beta c_2^2}I_m
		\end{bmatrix}.
	\end{aligned}
\end{equation*}
Then it is not difficult to obtain that
\begin{equation*}
	\Theta^\top \text{diag}\left(\begin{bmatrix}
		0& P \\ P& 0
	\end{bmatrix},\nu\begin{bmatrix}
		c^{-1}& 0 \\ 0& -P
	\end{bmatrix},\tau D_r^{-1}\right)\Theta\succ 0,
\end{equation*} where
\begin{equation}
	\Theta^\top=
	\begin{bmatrix}
		\begin{matrix}
			A_K& -I_{N-1} \\ K^\top& 0 \\ I_{N-1}& 0 \\ K^\top& 0
		\end{matrix}\left|
		\begin{matrix}
			0& B_1 \\ -I_m& 0 \\ 0& 0 \\ 0& 0
		\end{matrix}
		\right| \begin{matrix}
			I_{N-1}& 0& 0\\ 0& 0& 0\\ 0& -I_{N-1}& 0\\ 0& 0& -I_m
		\end{matrix}
	\end{bmatrix}
\end{equation}
and
\begin{equation}
	D_r=
	\begin{bmatrix}
		-I_{N-1}& 0& 0 \\ 0& \alpha c_1^2I_{N-1}& 0 \\ 0& 0& \beta c_2^2I_m
	\end{bmatrix}
\end{equation}
With the dualization lemma \cite{Herrmann2007}, we obtain
\begin{equation*}
	\tilde{\Theta}^\top=
	\begin{bmatrix}
		\begin{matrix}
			I_{N-1}& A_K^\top \\ 0& B_1^\top \\ 0& I_{N-1}
		\end{matrix}\left|
		\begin{matrix}
			K& 0 \\ 0& I_{N-1} \\ 0& 0
		\end{matrix}
		\right| \begin{matrix}
			0& I_{N-1}& K \\ 0& 0& 0 \\ I_{N-1}& 0& 0
		\end{matrix}
	\end{bmatrix},
\end{equation*}
and the negative definiteness of 
\begin{equation}\label{sum of three terms}
	\tilde{\Theta}^\top \text{diag}\left(\begin{bmatrix}
		0& P^{-1} \\ P^{-1}& 0
	\end{bmatrix},\frac{1}{\nu}\begin{bmatrix}
		c& 0 \\ 0& -P^{-1}
	\end{bmatrix},\frac{1}{\tau}D_r\right)\tilde{\Theta}.
\end{equation}
The above matrix \eqref{sum of three terms} can be written as the sum of three terms. Multiply it by $[z^\top \ (K^\top z)z^\top \ r(z,u)^\top]$ from the left and the transpose from the right, then the second term becomes
\begin{equation}
	\nu^{-1}\left(K^\top z\right)^2\left(c-z^\top P^{-1}z\right)\geq 0
\end{equation}
from the geometry of $\mathbb{X}_{RoA}$ as well as the third term satisfies
\begin{equation}
	\begin{aligned}
		&\tau^{-1}\left[\alpha c_1^2\|z\|^2+\beta c_2^2\|u\|^2-\|\hat{r}(z,u)\|^2\right] \\
		\geq &\tau^{-1}\left[\left(c_1\|z\|+c_2 \|u\|\right)^2-\|\hat{r}(z,u)\|^2\right] \geq 0
	\end{aligned}
\end{equation}
with probability $1-\delta$ according to Proposition 1, which is due to $\frac{1}{\alpha}+ \frac{1}{\beta}=1\Rightarrow\alpha\beta=\alpha+\beta$ and
\begin{equation*}
	\begin{aligned}
		&(\alpha-1)c_1^2\|z\|^2+(\beta-1)c_2^2 \|u\|^2 \\
		\geq&\sqrt{(\alpha-1)(\beta-1)}c_1c_2\|z\|\|=c_1c_2\|z\|\|u\|.
	\end{aligned}
\end{equation*}
Then we obtain the negatively definite property of time derivative of $V(x)$ from the first term after multiplication, i.e.,
\begin{equation}\label{negative Lyapunov derivative}
	\frac{\mathrm{d}}{\mathrm{d}t}V(x)=z^\top P^{-1}\left(\frac{\mathrm{d}z}{\mathrm{d}t}\right)+\left(\frac{\mathrm{d}z}{\mathrm{d}t}\right)^\top P^{-1}z<0
\end{equation}
which shows the forward invariance of $\mathbb{X}_{RoA}$.

Since the dictionary function $\Psi(x)$ is continuously differentiable, it is also Lipschitz continuous, i.e., $\exists L_z>0,\ \text{s.t.} \|z\|=\|\Psi(x)\|\leq L_z\|x\|$, which leads to
\begin{equation}
	V(x)\leq \lambda_{\max}(P^{-1})\|z\|^2\leq \lambda_{\max}(P^{-1})L_z^2\|x\|^2.
\end{equation}
Furthermore, since there exists a matrix $C\in\mathbb{R}^{n\times N}$ such that $x=Cz=C\Psi(x)\Rightarrow\|x\|\leq\|C\|\|z\|$, then
\begin{equation}
	V(x)\geq \lambda_{\min}(P^{-1})\|z\|^2\geq \frac{\lambda_{\min}(P^{-1})}{\|C\|^2}\|x\|^2.
\end{equation}
Together with \eqref{negative Lyapunov derivative}, we conclude the closed-loop exponential stability.
\qed

\end{document}